\documentclass[twocolumn,showpacs,aps,prd]{revtex4-1}
\usepackage{epsfig}
\usepackage{graphicx}
\usepackage{dcolumn}
\usepackage{bm}
\usepackage{ltablex,booktabs}
\usepackage{overpic}
\usepackage{subfigure}
\usepackage{float}
\usepackage{color}
\usepackage{amsmath}
\usepackage{mathcomp}
\usepackage{mathrsfs}
\usepackage{multirow}
\usepackage{rotating}
\usepackage{amssymb}
\usepackage{gensymb}
\usepackage{amsmath}
\usepackage{tabularx}
\usepackage{tikz-feynman}
\usepackage{lineno}
\usepackage[bookmarksnumbered, pdfstartview=FitH,colorlinks,urlcolor=blue, citecolor=blue,linkcolor=blue] {hyperref}

\begin{document}
\title{Study of $D_{s}^{+} \rightarrow f_{0}(980)\rho^+$ and $\phi \pi^+$ decays through $D_{s}^{+}\rightarrow \pi^{+}\pi^{+}\pi^{-}\pi^{0}$}
\author{
\begin{small}
\begin{center}
M.~Ablikim$^{1}$, M.~N.~Achasov$^{4,c}$, P.~Adlarson$^{75}$, O.~Afedulidis$^{3}$, X.~C.~Ai$^{80}$, R.~Aliberti$^{35}$, A.~Amoroso$^{74A,74C}$, Q.~An$^{71,58,a}$, Y.~Bai$^{57}$, O.~Bakina$^{36}$
% ...
\\
\vspace{0.2cm}
(BESIII Collaboration)\\
\vspace{0.2cm} {\it
$^{1}$ Institute of High Energy Physics, Beijing 100049, People's Republic of China\\
% ...
$^{3}$ Bochum  Ruhr-University, D-44780 Bochum, Germany\\
$^{4}$ Budker Institute of Nuclear Physics SB RAS (BINP), Novosibirsk 630090, Russia\\
% ...
$^{35}$ Johannes Gutenberg University of Mainz, Johann-Joachim-Becher-Weg 45, D-55099 Mainz, Germany\\
$^{36}$ Joint Institute for Nuclear Research, 141980 Dubna, Moscow region, Russia\\
% ...
$^{57}$ Southeast University, Nanjing 211100, People's Republic of China\\
$^{58}$ State Key Laboratory of Particle Detection and Electronics, Beijing 100049, Hefei 230026, People's Republic of China\\
% ...
$^{71}$ University of Science and Technology of China, Hefei 230026, People's Republic of China\\
% ...
$^{74}$ University of Turin and INFN, (A)University of Turin, I-10125, Turin, Italy; (B)University of Eastern Piedmont, I-15121, Alessandria, Italy; (C)INFN, I-10125, Turin, Italy\\
$^{75}$ Uppsala University, Box 516, SE-75120 Uppsala, Sweden\\
% ...
$^{80}$ Zhengzhou University, Zhengzhou 450001, People's Republic of China\\
\vspace{0.2cm}
$^{a}$ Deceased\\
% ...
$^{c}$ Also at the Novosibirsk State University, Novosibirsk, 630090, Russia\\
% ...
}      
\end{center}
\vspace{0.4cm}
\end{small}
}
\noaffiliation{}

\begin{abstract}
We perform the first amplitude
analysis of $D^+_s \to \pi^+\pi^+\pi^-\pi^0$ decays, based on 
data samples of electron-positron collisions recorded with the
BESIII detector at center-of-mass energies between 4.128 and 4.226~GeV,
corresponding to an integrated luminosity of 7.33~fb$^{-1}$.   We report the
observation of $D_{s}^{+} \to f_0(980)\rho(770)^{+}$ with a statistical significance
greater than 10$\sigma$ and determine the branching fractions
$\mathcal{B}(D_s^+\to\pi^+\pi^+\pi^-\pi^0|_{{\rm non}-\eta})=(2.04\pm0.08_{\rm stat.}\pm0.05_{\rm syst.})\%$
and $\mathcal{B}(D_s^+\to\eta\pi^+)=(1.56\pm0.09_{\rm stat.}\pm0.04_{\rm syst.})\%$.
Moreover, we measure the relative branching fraction between
$\phi\to\pi^+\pi^-\pi^0$ and $\phi\to K^+K^-$ to be
$\frac{\mathcal{B}(\phi(1020) \to \pi^+\pi^-\pi^0)}{\mathcal{B}(\phi(1020) \to K^+K^-)}=0.230 \pm 0.014_{\rm stat.} \pm 0.010_{\rm syst.}$,
which deviates from the world average value by more than $4\sigma$.
\end{abstract}

\maketitle

\linenumbers

The exploration of charmed-meson $D_{(s)}$ hadronic decays allows the
interplay of short-distance weak-decay matrix elements and long-distance
Quantum Chromodynamics~(QCD) interactions to be studied.  Moreover, measurements of the branching
fractions~(BFs) of charmed mesons can provide valuable insights for understanding the
amplitudes and phases induced by the strong
force~\cite{PDG, PRD79-034016, PRD81-074021, PRD84-074019}.
The amplitudes describing the weak decays of charmed mesons are dominated by
two-body processes, i.e.~$D_{(s)}\to VP$, $D_{(s)}\to PP$, $D_{(s)}\to SP$ and
$D_{(s)}\to VV$ decays, where $V$, $S$, and $P$ denote vector, scalar and
pseudoscalar mesons, respectively.
Significant progress has been achieved through a series of amplitude analyses on
hadronic charmed meson
decays~\cite{PDG, BESIII:2022kbq, BESIII:2019ly, LHCb:2018mzv,LHCb:2018xff}. However, there have been fewer studies of $D_{(s)}\to SV$ decays~\cite{PDG}, which means that the theoretical understanding of this process in less advanced, compared to other types of two-body decays. 
Among $D_{(s)}\to SV$ decays, $D^+_s \to f_0(980)\rho^+$
is of particular importance as it mainly involves a $W$-external-emission channel,
the BF of which can be precisely calculated in the absence of final-state
interactions, such as quark exchange or resonance
formation~\cite{BCKa03, BCKa0, BCKa02, Zhang:2022xpf}.
Final-state interactions are key ingredients in the production of light scalar
mesons, i.e. $f_0(500)$, $f_0(980)$, and $a_0(980)$, which are of particular interest
given the lack of consensus on whether these particles  are members of the 
normal scalar meson nonet or four-quark states.
In addition, the BESIII collaboration recently observed  abnormally large
BFs for the $D_s^+\to a_0(980)^{0(+)}\pi^{+(0)}$~\cite{BESIII:2019ly} and
$D_s^+\to a_0(980)^{0(+)}\rho^{+(0)}$~\cite{prd-104-L071101} decays, which could potentially be explained by
final-state rescattering effects~\cite{BCKa0, BCKa02}. 
Therefore, studying $D^+_s \to f_0(980)\rho^+$ through an amplitude
analysis of $D^+_s \to \pi^+\pi^+\pi^-\pi^0$ decays can experimentally constrain the
contribution from final-state interactions and help in understanding of the nature of the $f_0(980)$ meson.

Another interesting intermediate decay $D^+_s \to \phi \pi$ can be studied through $\phi \to \pi^+\pi^-\pi^0$. As the key reference channel for $D_s^+$
decays, $D_{s}^{+} \to \phi\pi^+$ is typically measured through
$\phi \to K^+K^-$~\cite{PDG}. However, studies of $\phi$ decays have primarily
been conducted in $e^+e^-$ annihilation and $K-p$ scattering
experiments~\cite{PDG, Parrour:1975rt, Mattiuzzi:1995eze, Dolinsky:1991vq, Akhmetshin:1995vz, Akhmetshin:1998se},
which often encounter challenges from complex backgrounds and various
interferences. The measurement of the BF of
$D_{s}^{+} \to \phi(\to \pi^+\pi^-\pi^0)\pi^+$, along with
$\mathcal{B}(D_{s}^{+} \to \phi(\to K^+K^-)\pi^+)$~\cite{BESIII:2020ctr}, can
provide a new method to determine the relative BF of
$R_\phi = \mathcal{B}(\phi\to \pi^+\pi^-\pi^0)/\mathcal{B}(\phi\to K^+K^-)$ in a more
controlled environment. Precise measurements of the BFs of $\phi$ decays are crucial not only for studying the strong interaction~\cite{Yukawa:1935xg,STAR:2022fan} but also for investigating $B$ decays which involve $\phi$ mesons~\cite{LHCb:2023exl,CDF:2005apk,BaBar:2003mjy,BaBar:2006ahy}.

Finally, the decay $D_{s}^{+} \to \omega\pi^+$
with $\omega \to \pi^+\pi^-\pi^0$, occurs solely via the
$W$-annihilation process. A precise measurement of its BF can help improve the theoretical understanding,
as current calculations suffer from large uncertainties~\cite{Cheng:2012wr, Qin:2013tje, Cheng:2010ry, Cheng:2016ejf}. The
BESIII Collaboration has reported the BF of this decay to be
$\mathcal{B}(D_{s}^{+} \to \omega\pi^+)=(1.77\pm0.32_{\rm stat.}\pm0.13_{\rm syst.})\times 10^{-3}$~\cite{PRD99-091101}.
In this Letter, we provide a more  precise measurement of the BF using
a larger data set through amplitude analysis, which takes the interference effect with other $D^+_s \to \pi^+\pi^+\pi^-\pi^0$ processes into account. In addition, the $D_{s}^{+} \to \pi^+\pi^+\pi^-\pi^0$ decay also contains a rich system of other
possible  intermediate components, such as
$D_{s}^{+} \to \eta \pi^{+}$, $D_{s}^{+} \to f_0(500)\rho^{+}$,
$D_{s}^{+} \to f_0(1370)\rho^{+}$, $D_{s}^{+} \to f_2(1270)\rho^{+}$,
$D_{s}^{+} \to \rho^0\rho^{+}$, $D_{s}^{+} \to a^+_1\pi^{0}$, etc. Studying the
relative contributions of these intermediate resonances can benefit the
understanding of the strong interaction at low energies and the $D_s^+$ weak-decay mechanism.

In this Letter, we present the first amplitude analysis of the decay
$D_{s}^{+} \to \pi^+\pi^+\pi^-\pi^{0}$ using 7.33~$\rm fb^{-1}$ of $e^+e^-$ collision data
collected with the BESIII detector at center-of-mass energies between 4.128 and
4.226~GeV. At these energies, $D_{s}^{*\pm}D_s^{\mp}$ events provide an ideal environment for the study of
$D_{s}^{+}$ physics. Throughout this Letter, charge-conjugated modes and exchange symmetry of two
identical $\pi^+$ are implied. The resonances $\phi(1020)$, $\omega(782)$, $\rho(770)^{+/0}$, and $a_1(1260)^{+/0}$ are referred to as $\phi$, $\omega$, $\rho^{+/0}$, and $a_1^{+/0}$, respectively.

The BESIII detector~\cite{Ablikim:2009aa} records symmetric $e^+e^-$ collisions
provided by the BEPCII storage ring~\cite{Yu:IPAC2016-TUYA01} in the
center-of-mass energy range from 1.85 to 4.95~GeV~\cite{Ablikim:2019hff}. Large
samples of Monte Carlo~(MC) simulated events are produced with
{\sc geant4}-based~\cite{GEANT4} software, and are used to determine the
detection efficiency and to estimate the background contributions. The
beam-energy spread and initial-state radiation in the $e^+e^-$ annihilation
are modeled with the generator {\sc kkmc}~\cite{KKMC}. Inclusive MC samples
of 40 times the size of the data sample are used to simulate the background
contributions.  The inclusive MC sample includes the production of open charm
processes, the ISR production of vector charmonium(-like) states,
and the continuum processes incorporated in {\sc kkmc}.
All particle decays are modeled with
{\sc evtgen}~\cite{ref:evtgen} using BFs either taken from the
Particle Data Group~\cite{PDG}, when available, or otherwise estimated with
{\sc lundcharm}~\cite{ref:lundcharm}. Final-state radiation from charged particles is incorporated
using the {\sc photos} package~\cite{photos}. The signal detection efficiencies and signal
shapes are obtained from signal MC samples, in which the signal
$D_s^+ \to \pi^+ \pi^+ \pi^- \pi^0$ decay is simulated using the model from
the amplitude analysis introduced in this Letter.

Signal events are from the
$e^{+}e^{-} \to D_{s}^{*+}D_{s}^{-}+c.c.\to \gamma D_{s}^{+}D_{s}^{-}$ process, where
$D_{s}^{*+(-)}D_{s}^{-(+)}$ are produced without additional hadronic particles,
which provides a clean environment for amplitude analysis and precise
measurement of the absolute BFs of $D^{\pm}_{s}$ hadronic decays. We utilize a
double-tag (DT) technique~\cite{MarkIII-tag,Ke:2023qzc,Li:2021iwf} to study the signal process.  In this procedure 
there are two types of samples: single-tag (ST) events, which are reconstructed with a  $D^-_s$ tag; and DT events, which are reconstructed with both a $D^-_s$ tag and signal $D^+_s$. In this analysis, the ST tag $D^-_s$ candidates are reconstructed through
seven modes: $D_{s}^{-} \to K_{S}^{0}K^{-}$, $D_{s}^{-} \to K^{+}K^{-}\pi^{-}$,
$D_{s}^{-} \to K^{+}K^{-}\pi^{-}\pi^{0}$,
$D_{s}^{-} \to K_{S}^{0}K^{+}\pi^{-}\pi^{-}$,
$D_{s}^{-} \to \pi^{-}\eta_{\gamma\gamma}$,
$D_{s}^{-} \to \pi^{-}\eta'_{\pi^+\pi^-\eta_{\gamma\gamma}}$, and
$D_{s}^{-} \to K^-\pi^-\pi^+$. Here, the $K_{S}^{0}$, $\pi^{0}$, $\eta$, and
$\eta'$ mesons are reconstructed from
$K_{S}^{0} \to \pi^{+}\pi^{-}$, $\pi^{0} \to \gamma\gamma$,
$\eta \to \gamma\gamma$, and $\eta' \to \pi^{+}\pi^{-}\eta$ decays,
respectively. The selection criteria for charged and neutral particle
candidates are identical to those used in Ref.~\cite{prd-104-L071101}. For the decay mode
$D_{s}^{-} \to K^{+}K^{-}\pi^{-}\pi^{0}$, we reject events with $K^+K^-$ invariant mass
above 1.05~GeV/$c^{2}$ to suppress background. The DT candidates are selected by
reconstructing the signal process $D_{s}^{+} \to \pi^{+}\pi^{+}\pi^{-}\pi^{0}$
from the remaining particles that are not used in the ST reconstruction.

The invariant masses of the ST and DT $D_{s}^{\pm}$ candidates, denoted 
$M_{\text{tag}}$ and $M_{\text{sig}}$, respectively, are required to be within the range
\mbox{[1.87, 2.06]}~GeV$/c^{2}$. We calculate the recoiling mass
$M_{\text{rec}}=\{[E_{\text{cm}}-(|\vec{p}_{D_{s}^{-}}|^{2}+m^{2}_{D_{s}^{-}})^{1/2}]^{2}-|\vec{p}_{D_{s}^{-}}|^{2}\}^{1/2}$
in the $e^+e^-$ center-of-mass system, where $E_{\text{cm}}$ is the center-of-mass energy of the data sample, $\vec{p}_{D_{s}^{-}}$ is the
momentum of the reconstructed $D_{s}^{-}$ and $m_{D_{s}^{-}}$ is the known mass
of the $D_{s}^{-}$ meson~\cite{PDG}. The value of $M_{\text{rec}}$ is required
to be in the range \mbox{[2.05, 2.18]}~GeV$/c^{2}$ for the data sample collected at
4.178~GeV to suppress the non-$D_{s}^{*\pm}D_{s}^{\mp}$ events. The
$M_{\text{rec}}$ ranges for the other data samples are the same as those in Ref.~\cite{prd-104-L071101}.

To suppress background from $D^{+}_{s} \to K^{0}_{S}\pi^{+}\pi^{0}$ decays, events are
rejected if any of the two $\pi^{+}\pi^{-}$ combinations of the candidate signal decay 
has an invariant mass lying within the range \mbox{[0.46, 0.52]}~GeV$/c^{2}$.
The decay $D^{+}_{s} \to \eta \pi^{+}$ is also considered as background because $\eta \to \pi^+\pi^-\pi^0$ lies at the boundary of the phase
space and thus has little interference with other intermediate decays in the $D^{+}_{s} \to \pi^{+}\pi^{+}\pi^{-}\pi^{0}$ process. Therefore, events are rejected if any of the two $\pi^+\pi^-\pi^0$ combinations in the final $\pi^+\pi^+\pi^-\pi^0$ state has an invariant mass within the $\eta$ mass range of \mbox{[0.52, 0.58]}~GeV$/c^{2}$.

To reduce combinatorial background, a seven-constraint~(7C) kinematic fit~\cite{Yan:2010zze} is applied to the
$e^+e^- \to D^{*\pm}_{s}D^{\mp}_{s} \to \gamma D^{+}_{s}D^{-}_{s}$ candidates,
where $D^{-}_{s}$ decays to one of the tag modes and $D^{+}_{s}$ decays to the
signal mode. The constraints are: four-momentum conservation in the center-of-mass system, and imposing that the invariant mass of
$\pi^{0}$ from the signal decay, the reconstructed $D_{s}^{-}$ from the tag decays, and the $D^{*+}_{s}$ candidate have their 
PDG values~\cite{PDG}. If there are multiple candidate combinations, the
combination with the minimum $\chi^{2}$ of the 7C kinematic fit is retained.

An observable,
$M_{\text{rec0}}=\{[E_{\text{cm}}-(\vec{p}^{2}_{D^{+}_{s}\gamma}+m^{2}_{D^{*\pm}_{s}})^{1/2}]^{2}-|\vec{p}_{D^{+}_{s}}|^{2}\}^{1/2}$
, is required to lie within the range \mbox{[1.958, 1.986]}~GeV$/c^{2}$. The energy of the
radiative photon from the $D^{*\pm}_{s}$ is required to be less than 0.18~GeV. The invariant
mass of the $D^{*\pm}_{s}$ candidate must be within \mbox{[2.066, 2.135]}~GeV$/c^{2}$.
Finally, the mass of the signal $D^{+}_{s}$ candidate is required to be within the range
\mbox{[1.930, 1.985]}~GeV$/c^{2}$.

In particular for amplitude analysis, to achieve a better resolution for the reconstructed momentum, an additional
constraint is added, imposing that the reconstructed signal $D^{+}_{s}$ mass has the PDG value.
 The four momenta of candidate events are updated following this eight-constraint  (8C)
kinematic fit for the amplitude analysis.

The data sets are divided into four categories according to the center-of-mass energy range: 4.13-4.16, 4.178, 4.189-4.219, and
4.226~GeV.   We fit the $D_s^+$ peaks in these samples with a signal shape taken from MC simulation, convolved with a Gaussian function, and a shape for the background distribution also taken from simulation.   The purities are determined to be  $(83.8\pm1.1)\%$,
$(81.0\pm0.7)\%$, $(80.2\pm1.0)\%$, and $(75.7\pm2.2)\%$, with corresponding signal yields of $189\pm17$, $778\pm35$, $448\pm26$, and $137\pm15$, respectively.

A simultaneous unbinned maximum-likelihood fit is performed on the four categories of data. The probability density function (PDF) is constructed depending on the momenta of the four final-state particles, using a signal-background model: $\text{PDF}(\bm{x})=\xi f_{S}(\bm{x}) + (1-\xi) f_{B}(\bm{x})$, where $\xi$ is the purity of data set, $\bm{x}$ is the location in phase space of the decay (determined by the momenta of the four final particles), $f_{S}$ is the normalized signal-process distribution function, and $f_{B}$ is the normalized background-distribution function. The signal model is constructed as a coherent sum of intermediate processes $M(\bm{x})=\Sigma \rho_{n} e^{i\phi_{n}} A_{n}(\bm{x})$, where $\rho_{n} e^{i\phi_{n}}$ is the complex coefficient of the $n$-th amplitude. The component amplitude $A_{n}(\bm{x})$ is given by $A_{n}=P^{1}_{n}P^{2}_{n}S_{n}F^{1}_{n}F^{2}_{n}F^{3}_{n}$, where the indices 1, 2, and 3 correspond to the two subsequent intermediate resonances and the $D^{+}_{s}$ meson, $F^{i}_{n}$ the Blatt-Weisskopf barrier factor~\cite{Blatt-Weisskopf,Zou:2002ar}, and $P^{i}_{n}$ the propagator of the intermediate resonance. The function $S_{n}$ is the spin factor describing the \textit{L-S} coupling in the amplitude and is constructed using the covariant-tensor formalism~\cite{Zou:2002ar,Jing:2023co,Li:2023sp}. The propagators employed in this analysis are as follows: a relativistic Breit-Wigner~\cite{Jackson:1964zd} function for the $f_{0}(1370)$, $f_{2}(1270)$, $\pi(1300)$, $a_1$, $\rho(1450)$, $\phi$, and $\omega$ resonance; a Gounaris-Sakurai~\cite{Gounaris:1968mw} line shape for the $\rho$ resonance; and a coupled Flatt$\acute{e}$~\cite{Flatte:1976xu} formula for the $f_{0}(980)$ resonance, whose parameters are taken from Refs.~\cite{BESIII:2016tqo,BES:2004twe}. The other physical parameters for involving resonances in amplitude analysis are listed in Table~\ref{tab:resonparam}

The background model $B(\bm{x})$ is constructed from inclusive MC samples by using a multidimensional kernel density estimator~\cite{kernel} with five independent Lorentz invariant variables ($M_{\pi^{+}\pi^{+}}$, $M_{\pi^{+}\pi^{-}}$, $M_{\pi^{+}\pi^{0}}$, $M_{\pi^{-}\pi^{0}}$, and $M_{\pi^{+}\pi^{-}\pi^{0}}$). The extracted shape shows good consistence with data side-band. As a consequence, the combined PDF can be written as
\begin{align}
  \epsilon R_{4} \left[ \xi \frac{|M(\bm{x})|^{2}}{\int \epsilon |M(\bm{x})|^{2} R_{4} {\rm d}\bm{x}} + (1-\xi) \frac{B_{\epsilon}(\bm{x})}{\int \epsilon B_{\epsilon}(\bm{x}) R_{4} {\rm d}\bm{x}} \right],
\label{eq:pdf}
\end{align}
where $\epsilon$ is the acceptance function determined with phase-space (PHSP) MC samples generated with a uniform distribution over final particles' momentum of $D^{+}_{s} \to \pi^{+}\pi^{+}\pi^{-}\pi^{0}$ decays, $B_{\epsilon}(\bm{x})$ is defined as $B(\bm{x})/\epsilon$, and $R_{4}{\rm d}\bm{x}$ is an element of four-body PHSP. The normalization integral in the denominator is calculated by the MC technique described in Ref.~\cite{CLEO:2012beo}.

\begin{figure}[htp]
  \begin{center}
    \includegraphics[width=0.45\textwidth]{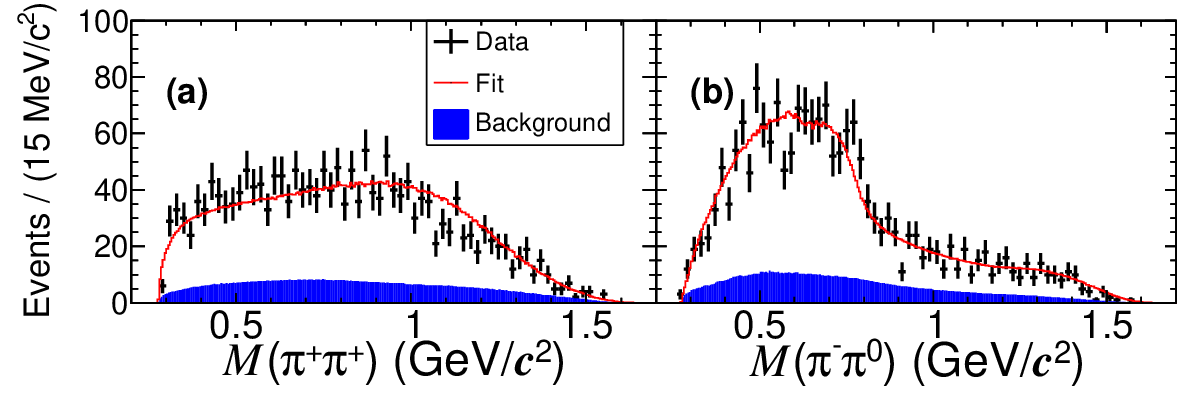}
    \includegraphics[width=0.45\textwidth]{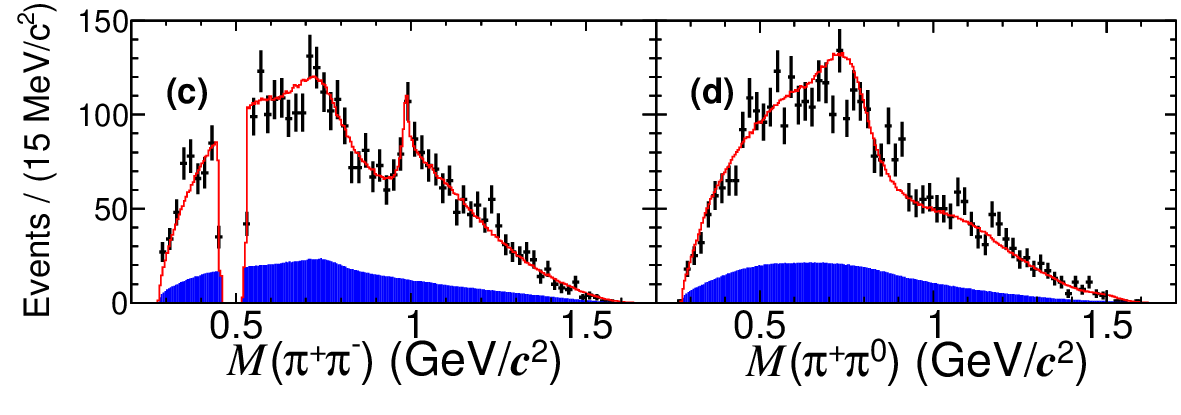}
    \includegraphics[width=0.45\textwidth]{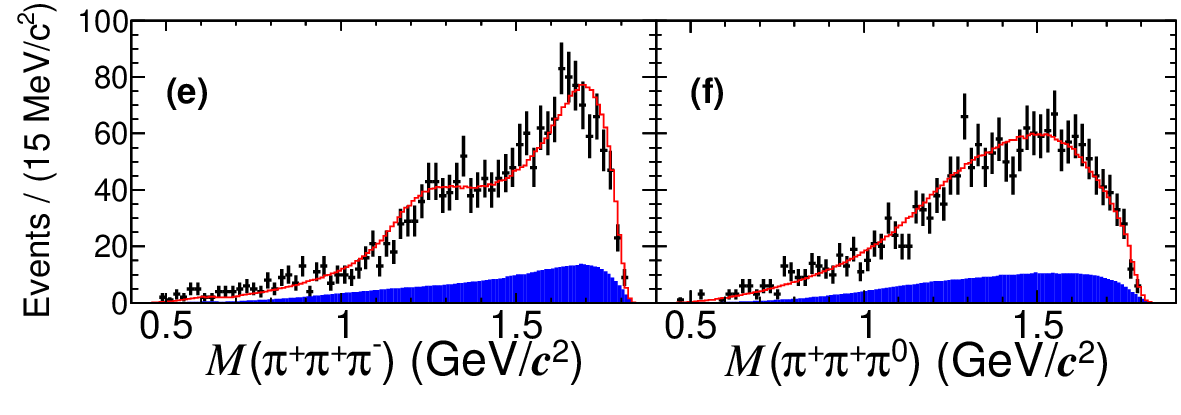}
    \includegraphics[width=0.23\textwidth]{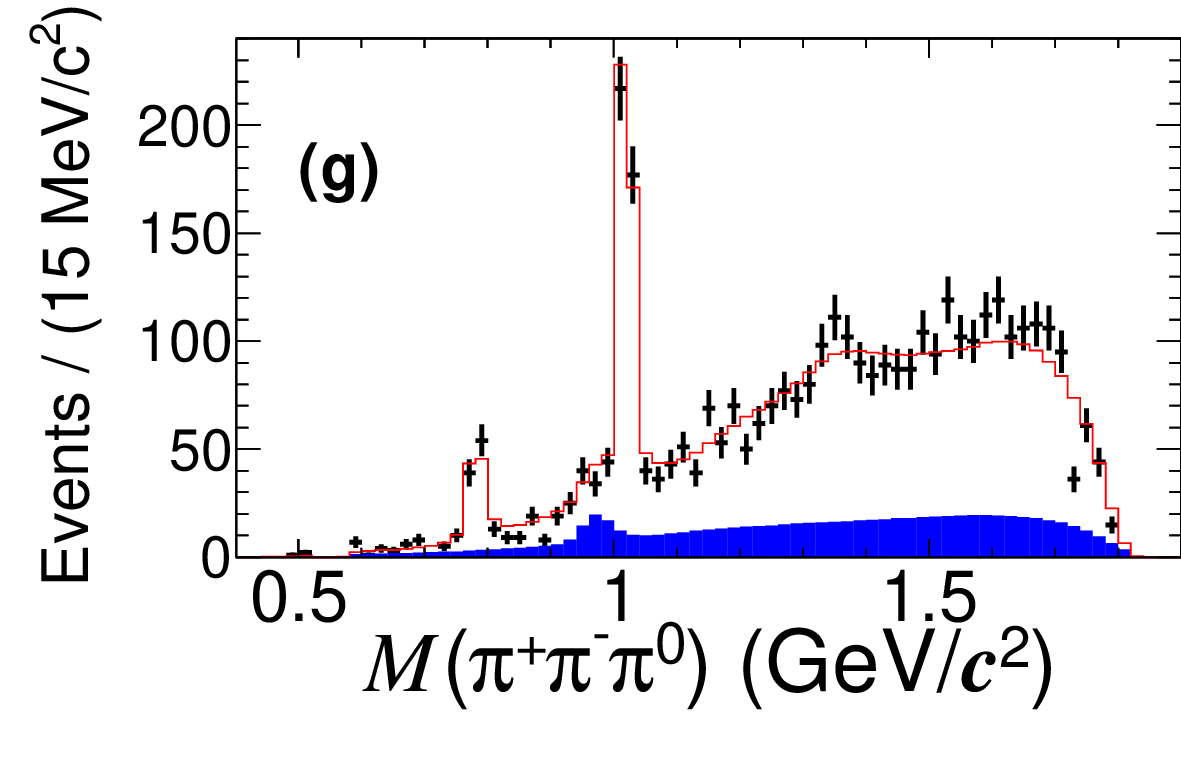}
    \caption{Projections on (a) $M_{\pi^+\pi^+}$, (b) $M_{\pi^-\pi^0}$, (c) $M_{\pi^+\pi^-}$,
      (d) $M_{\pi^+\pi^0}$, (e) $M_{\pi^+\pi^+\pi^-}$, (f)
      $M_{\pi^+\pi^+\pi^0}$, (g) $M_{\pi^+\pi^-\pi^0}$ of the amplitude analysis. 
The combinations of two identical
      $\pi^+$ are added in (c), (d), and (g), because of the exchange
      symmetry.}
    \label{fig:ampfit}
  \end{center}
\end{figure}

In the amplitude analysis, the initial model is constructed from those significant components known to be present, namely $\phi\pi^{+}$, $\omega\pi^{+}$, $f_{0}(980)\rho^{+}$ , and $f_{0}(1370)\rho^{+}$. Then, further components are added, one at a time, to the fit. The statistical significance of a component is calculated from the resulting change of likelihood and number of degrees of freedom. Only those components with significance larger than $5\sigma$ are retained for the optimal model.  Some components, such as non-resonance, $f_0(1500)$ and $a_2(1320)$, because of limited siginificance, are not included in the model. The dominant Cabibbo-favored process $D^{+}_{s} \to f_{0}(1370)\rho^{+}$ is selected as the reference component, with its phase fixed to zero and magnitude to unity. 
The coefficients of the isospin-related sub-decays of the  $\phi$, $\omega$, and $a_1$ are related by Clebsch-Gordan coefficients. The final model contains eleven components, as listed in Table~\ref{tab:amptab}. The mass projections of the fit are shown in Fig.~\ref{fig:ampfit}, and the angular distribution of final state $\pi$ mesons can be found in Ref.~\cite{Supplement}. The contribution of  the $n^{\text{th}}$ component relative to the total BF is quantified by the fit fraction~(FF) defined as $\text{FF}_{n}=\int |\rho_{n} A_{n}(\bm{x})|^{2} R_{4} {\rm d}\bm{x} / \int |M(\bm{x})|^{2} R_{4} {\rm d}\bm{x}$. The measured phases and FFs for the different components in the optimal fit are listed in Table~\ref{tab:amptab}. Dominant interference is found to be from $f_0(980)$. The mass projections for contributions from different components, and the interference from $f_0(980)$ can be found at Ref.~\cite{Supplement}.

We determine the systematic uncertainties by taking the differences between the values of $\phi_{n}$ and $\text{FF}_{n}$ found by the optimal fit and those found from fit variations. The masses and widths of intermediate states, such as $f_0(1370)$, $\rho(1450)$ and $\pi(1300)$, are varied by $\pm 1 \sigma$~\cite{PDG}. The masses and coupling constants of the $f_{0}(980)$ are varied within the uncertainties reported in Refs.~\cite{BESIII:2016tqo,BES:2004twe}. The barrier radii for $D^{+}_{s}$ and the other intermediate states are varied by $\pm 1~\text{GeV}^{-1}$. The uncertainties from detector effects are investigated by weighting PHSP MC samples according to data-MC difference. The same method is also employed in Ref.~\cite{BESIII:2020ctr}. The uncertainty related to background is estimated by varying the estimated purity within its statistical uncertainty. The total uncertainties are obtained by adding the separate contributions in quadrature, as listed in Table~\ref{tab:amptab}.

\begin{table*}[htbp]
  \renewcommand\arraystretch{1.25}
  \setlength{\tabcolsep}{7pt}
  \caption{Physical parameters for the resonances involved in amplitude analysis.
  }\label{tab:resonparam}
  \begin{center}
    \begin{tabular}{lccc}
      \hline \hline
      Resonance & Mass~(MeV$/c^{2}$) & Width~(MeV$/c^{2}$) \\
      \hline
      $f_0(1370)$ & 1350 & 265 \\
      $f_0(980)$ & 960 & $g_1=165,g_2=4.21g_1$~\cite{BESIII:2016tqo,BES:2004twe} \\
      $f_2(1270)$ & 1275 & 190 \\
      $\rho(770)$ & 775 & 149 \\
      $\rho(1450)$ & 1465 & 400 \\
      $\phi(1020)$ & 1019 & 4 \\
      $\omega(782)$ & 783 & 9 \\
      $a_1(1260)$ & 1230 & 420 \\
      $\pi(1300)$ & 1371 & 314 \\
      \hline \hline
    \end{tabular}
  \end{center}
\end{table*}

\begin{table*}[htbp]
  \renewcommand\arraystretch{1.25}
  \setlength{\tabcolsep}{7pt}
  \caption{Phases, FFs, and BFs for various intermediate processes in $D^+_s \to \pi^+\pi^+\pi^-\pi^0$ decay. The first and the
    second uncertainties are statistical and systematic, respectively. The subsequent decay is given in parentheses, with the subscript \textit{S} and \textit{P} indicating 
    the spatial wave mode.
  }\label{tab:amptab}
  \begin{center}
    \begin{tabular}{lccc}
      \hline \hline
      Component & Phase~(rad) & FF~(\%) & BF~($10^{-3}$) \\
      \hline
      $f_{0}(1370)\rho^{+}$ & 0.0(fixed) & $24.9\pm3.8\pm2.1$ & $5.08\pm0.80\pm0.43$ \\
      $f_{0}(980)\rho^{+}$ & $3.99\pm0.13\pm0.07$ & $12.6\pm2.1\pm1.0$ & $2.57\pm0.44\pm0.20$ \\
      $f_{2}(1270)\rho^{+}$ & $1.11\pm0.10\pm0.10$ & $\hspace{0.17cm} 9.5\pm1.7\pm0.6$ & $1.94\pm0.36\pm0.12$ \\
      $(\rho^{+}\rho^{0})_{S}$ & $1.10\pm0.18\pm0.10$ & $\hspace{0.17cm} 3.5\pm1.2\pm0.6$ & $0.71\pm0.25\pm0.12$ \\
      $(\rho(1450)^{+}\rho^{0})_{S}$ & $0.43\pm0.18\pm0.17$ & $\hspace{0.17cm} 4.6\pm1.3\pm0.8$ & $0.94\pm0.27\pm0.16$ \\
      $(\rho^{+}\rho(1450)^{0})_{P}$ & $4.58\pm0.16\pm0.09$ & $\hspace{0.17cm} 8.6\pm1.3\pm0.4$ & $1.75\pm0.27\pm0.08$ \\
      $\phi((\rho\pi)\to\pi^+\pi^-\pi^0)\pi^{+}$ & $2.90\pm0.15\pm0.18$ & $24.9\pm1.2\pm0.4$ & $5.08\pm0.32\pm0.10$ \\
      $\omega((\rho\pi)\to\pi^+\pi^-\pi^0)\pi^{+}$ & $3.22\pm0.21\pm0.09$ & $\hspace{0.17cm} 6.9\pm0.8\pm0.3$ & $1.41\pm0.17\pm0.06$ \\
      $a^{+}_{1}(\rho^{0}\pi^{+})_{S}\pi^{0}$ & $3.78\pm0.16\pm0.12$ & $12.5\pm1.6\pm1.0$ & $2.55\pm0.34\pm0.20$ \\
      $a^{0}_{1}((\rho\pi)_{S}\to\pi^+\pi^-\pi^0)\pi^{+}$ & $4.82\pm0.15\pm0.12$ & $\hspace{0.17cm} 6.3\pm1.9\pm1.2$ & $1.29\pm0.39\pm0.24$ \\
      $\pi(1300)^{0}((\rho\pi)_{P}\to\pi^+\pi^-\pi^0)\pi^{+}$ & $2.22\pm0.14\pm0.08$ & $11.7\pm2.3\pm2.2$ & $2.39\pm0.48\pm0.45$ \\
      \hline \hline
    \end{tabular}
  \end{center}
\end{table*}

\begin{figure}[htp]
  \begin{center}
    \includegraphics[width=0.22\textwidth]{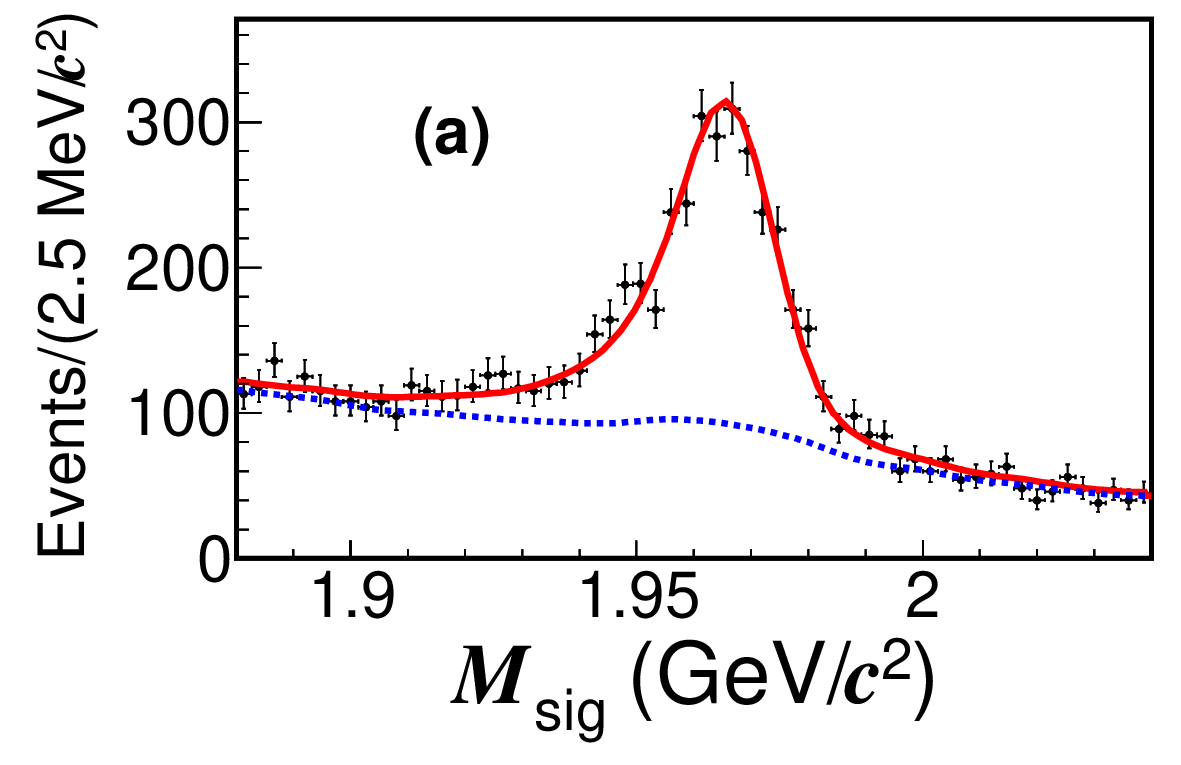}
    \includegraphics[width=0.22\textwidth]{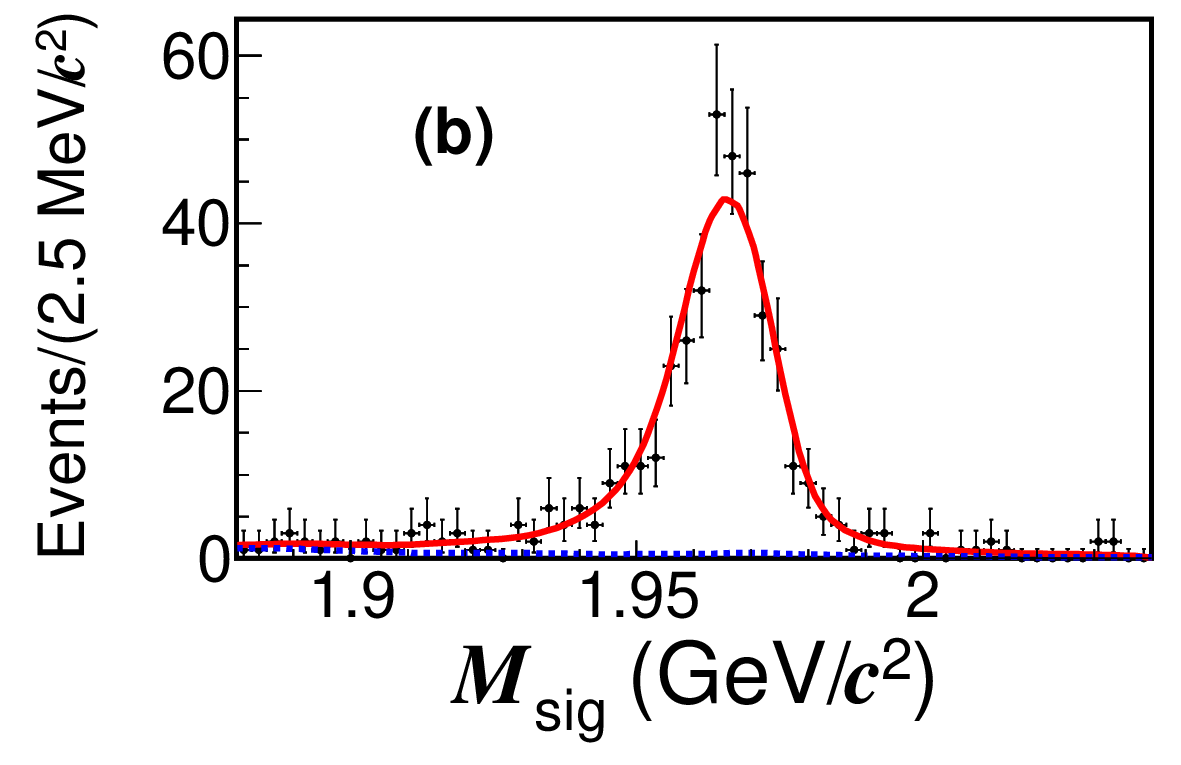}
    \caption{Fits to the $M_{\text{sig}}$ distributions of the DT candidates for (a) ``non-$\eta$" and (b) ``$\eta\pi$" contributions. 
    The data with error bars represent data from all samples, while the red solid lines are the total fits to the data. The dashed blue lines indicate the fitted background shapes.}
    \label{fig:bf}
  \end{center}
\end{figure}

The measurement of the $D^{+}_{s} \to \pi^+\pi^+\pi^-\pi^0$ BF is performed using
a DT technique based on seven ST modes, the same as for the amplitude analysis. It is performed separately for ``non-$\eta$" and ``$\eta\pi$" contributions. The ``$\eta\pi$" events are defined as those with the invariant mass of
any of the two $\pi^+\pi^-\pi^0$ combinations in the final state of $\pi^+\pi^+\pi^-\pi^0$, within the $\eta$ mass range of \mbox{[0.52, 0.58]}~GeV$/c^{2}$, with all other
events classified in the ``non-$\eta$" category. If there are multiple tag $D^{-}_{s}$
candidates for each tag mode, then the one with $M_{\text{rec}}$ closest to the
known mass of $D^{*\pm}_{s}$~\cite{PDG} is retained. A DT candidate with
average mass $(M_{\text{sig}}+M_{\text{tag}})/2$ closest to the known mass of
$D^{+}_{s}$~\cite{PDG} is retained for each tag mode. The ST yields ($Y_{\text{tag}}$) and DT yield ($Y_{\text{sig}}$) in data are determined from fits to the $M_{\text{tag}}$ and $M_{\text{sig}}$ distributions, respectively. The ST fit results are the same as Refs.~\cite{prd-104-L071101,BESIII:2020ctr}. The DT fits are shown in Fig.~\ref{fig:bf}. The signal shape is modeled with the shape from MC simulation convolved with a Gaussian resolution function, and the background is estimated from the inclusive MC sample.

These fits result in a total ST yield of $Y_{\text{tag}}=471617\pm1733$. For the ``non-$\eta$" part, the signal yield is $Y_{\text{sig,non-}\eta}=2489\pm91$ and for the ``$\eta\pi$" part the signal yield is $Y_{\eta\pi^{+}}=392\pm22$. An updated inclusive MC sample based on our amplitude analysis results is used to determine the ST efficiencies ($\epsilon^i_{\text{ST}}$) and DT efficiencies ($\epsilon^i_{\text{DT}}$). Substituting these results into $\mathcal{B}(D^{+}_{s}\to\pi^+\pi^+\pi^-\pi^0|_{\text{non-}\eta})=Y_{\text{sig,non-}\eta}/(\mathcal{B}(\pi^0\to\gamma\gamma)\times \Sigma_{i,\alpha}Y^{i,\alpha}_{\text{tag}}\epsilon^{i,\alpha}_{\text{DT}}/\epsilon^{i,\alpha}_{\text{ST}})$ and $\mathcal{B}(D^{+}_{s}\to\eta(\to\pi^+\pi^-\pi^0)\pi^+)=Y_{\text{sig,}\eta\pi^{+}}/(\mathcal{B}(\pi^0\to\gamma\gamma)\times \Sigma_{i,\alpha}Y^{i,\alpha}_{\text{tag}}\epsilon^{i,\alpha}_{\text{DT}}/\epsilon^{i,\alpha}_{\text{ST}})$, where $i$ denotes the $i$th tag mode and $\alpha$ denotes the $\alpha$th center-of-mass energy point, we obtain $\mathcal{B}(D^{+}_{s}\to\pi^+\pi^+\pi^-\pi^0|_{\text{non}-\eta})=(2.04\pm0.08)\%$ and $\mathcal{B}(D^{+}_{s}\to\eta(\to\pi^+\pi^-\pi^0)\pi^+)=(3.58\pm0.21)\times10^{-3}$, where the uncertainties are statistical only.

The systematic uncertainties for the BF measurement are categorized in five sources: (a) uncertainty from the number of ST $D^{-}_{s}$ mesons, estimated by considering the statistical effect related to the ST background, (b) the DT background shape, estimated by changing to alternative background shapes, (c) the $\pi^{\pm}$ tracking (PID) efficiency and $\pi^{0}$ reconstruction, estimated by studying related control samples of $D^+_s \to K^+K^-K^+K^-$ and $D^+_s \to K^+K^-K^+K^-\pi^0$ decays, (d) MC sample size and model, estimated by studying the change in result when varying the signal-model parameters, and (e) the knowledge of the  BFs of $\mathcal{B}(\pi^{0}\to\gamma\gamma)$ and $\mathcal{B}(\eta\to\pi^+\pi^-\pi^0)$~\cite{PDG}. Adding all sources of uncertainties in quadrature gives a total of $2.4\%$ systematic uncertainty for $\mathcal{B}(D^{+}_{s}\to\pi^+\pi^+\pi^-\pi^0|_{\text{non}-\eta})$, and $1.6\%$ for $\mathcal{B}(D^{+}_{s}\to\eta(\to\pi^+\pi^-\pi^0)\pi^+)$.

In summary, we measure the absolute BFs
$\mathcal{B}(D^{+}_{s}\to\pi^+\pi^+\pi^-\pi^0|_{\text{non}-\eta})=(2.04\pm0.08_{\text{stat.}}\pm0.05_{\text{syst.}})\%$
for the first time, and
$\mathcal{B}(D^{+}_{s}\to\eta(\to \pi^+\pi^-\pi^0)\pi^+)=(3.58\pm0.21_{\text{stat.}}\pm0.06_{\text{syst.}})\times10^{-3}$.
Utilizing $\mathcal{B}(\eta \to \pi^+\pi^-\pi^0)$ quoted from the PDG~\cite{PDG},
the BF $\mathcal{B}(D^+_s \to \eta\pi^+)$ is determined to be
$(1.56\pm0.09_{\text{stat.}}\pm0.04_{\text{syst.}})\%$. Moreover, we perform the first amplitude analysis of
$D^{+}_{s}\to\pi^+\pi^+\pi^-\pi^0|_{\text{non}-\eta}$ and report the
observation of $D^+_s \to f_{0}(980)\rho^+$. The phases and FFs of the
significant intermediate processes are summarized in Table~\ref{tab:amptab}.
The BFs for the intermediate processes are calculated as
$B_{n}=\text{FF}_{n} \times \mathcal{B}(D^{+}_{s}\to\pi^+\pi^+\pi^-\pi^0|_{\text{non}-\eta})$, which are summarized in Table~\ref{tab:amptab}.
The $D^+_s \to f_{0}(1370)(\to\pi^+\pi^-)\rho^+(\to\pi^+\pi^0)$ and
$D^+_s \to \phi(\to\pi^+\pi^-\pi^0)\pi^+$ contributions dominate with BFs of
$(5.08\pm0.80_{\text{stat.}}\pm0.43_{\text{syst.}}) \times 10^{-3}$ and $(5.08\pm0.32_{\text{stat.}}\pm0.10_{\text{syst.}}) \times 10^{-3}$,
respectively. 
The BF of $D^+_s \to f_{0}(980)(\to\pi^+\pi^-)\rho^+(\to\pi^+\pi^0)$
is found to be $(2.57\pm0.44_{\text{stat.}}\pm0.20_{\text{syst.}}) \times 10^{-3}$, which is valuable input for improving
understanding of the nature of  the $f_{0}(980)$ meson. Taking the BF of
$D^+_s \to \phi(\to K^+K^-)\pi^+$ from Ref.~\cite{BESIII:2020ctr}, enables the relative BF between
$\phi$ decays into $\pi^+\pi^-\pi^0$ and $K^+K^-$  to be calculated.  The result of 
\mbox{$R_\phi=\frac{\mathcal{B}(\phi(1020) \to \pi^+\pi^-\pi^0)}{\mathcal{B}(\phi(1020) \to K^+K^-)}=0.230 \pm 0.014_{\text{stat.}} \pm 0.010_{\text{syst.}}$},
significantly deviates from the PDG value \mbox{$R^{\text{PDG}}_\phi=\frac{\mathcal{B}(\phi(1020) \to \pi^+\pi^-\pi^0)}{\mathcal{B}(\phi(1020) \to K^+K^-)}=0.313\pm0.010$} by more than $4\sigma$~\cite{PDG}. Further checks on the $\phi\pi^+$ BF can be accessed through Ref.~\cite{Supplement}. The data in the PDG on the OZI suppressed $\phi \to \rho \pi$ decays are mostly from $e^+e^-$ collisions on the $\phi$ peak~\cite{PDG}. The possible interference effect between $e^+e^- \to \gamma^{*} \to \rho \pi$ and $e^+e^- \to \phi \to \rho \pi$ may not be well considered, 
 and therefore, BESIII $D_s$ decay provides another test of this ratio, in which non such interference arises. 
This is the first measurement of $R_\phi$ in hadronic decays of charmed mesons, and the lower than expected value motivates further studies. The BF of the $W$-annihilation decay
$D^+_s \to \omega(\to\pi^+\pi^-\pi^0)\pi^+$ is determined to be
$(1.41\pm0.17_{\text{stat.}}\pm0.06_{\text{syst.}}) \times 10^{-3}$. This result is a factor two more precise than
previous measurements,  and obtained in a manner that takes full account of  interference with other intermediate processes decaying into   the same final state.
The significantly improved
precision will benefit investigations of the underlying dynamics for
non-perturbative $W$-annihilation amplitudes and allow for better predictions of the BFs
and direct \textit{CP} violation of decays involving $W$-annihilation~\cite{Cheng:2012wr, Qin:2013tje, Cheng:2010ry, Cheng:2016ejf}. 

\begin{acknowledgements}
\label{sec:acknowledgement}
\vspace{-0.4cm}
The BESIII Collaboration thanks the staff of BEPCII and the IHEP computing center for their strong support. This work is supported in part by National Key R\&D Program of China under Contracts Nos. 2023YFA1606000, 2020YFA0406400, 2020YFA0406300; National Natural Science Foundation of China (NSFC) under Contracts Nos. 11635010, 11735014, 11835012, 11875054, 11935015, 11935016, 11935018, 11961141012, 12025502, 12035009, 12035013, 12061131003, 12192260, 12192261, 12192262, 12192263, 12192264, 12192265, 12221005, 12225509, 12235017, 12042507, 123B2077, 12342502; the Chinese Academy of Sciences (CAS) Large-Scale Scientific Facility Program; the CAS Center for Excellence in Particle Physics (CCEPP); Joint Large-Scale Scientific Facility Funds of the NSFC and CAS under Contract Nos. U2032104, U1832207; CAS Key Research Program of Frontier Sciences under Contracts Nos. QYZDJ-SSW-SLH003, QYZDJ-SSW-SLH040; 100 Talents Program of CAS; The Institute of Nuclear and Particle Physics (INPAC) and Shanghai Key Laboratory for Particle Physics and Cosmology; European Union's Horizon 2020 research and innovation programme under Marie Sklodowska-Curie grant agreement under Contract No. 894790; German Research Foundation DFG under Contracts Nos. 455635585, Collaborative Research Center CRC 1044, FOR5327, GRK 2149; Istituto Nazionale di Fisica Nucleare, Italy; Ministry of Development of Turkey under Contract No. DPT2006K-120470; National Research Foundation of Korea under Contract No. NRF-2022R1A2C1092335; National Science and Technology fund of Mongolia; National Science Research and Innovation Fund (NSRF) via the Program Management Unit for Human Resources \& Institutional Development, Research and Innovation of Thailand under Contract No. B16F640076; Polish National Science Centre under Contract No. 2019/35/O/ST2/02907; The Swedish Research Council; U. S. Department of Energy under Contract No. DE-FG02-05ER41374. We also thank Dr. Xiang-Kun Dong from helmholtz-Institut f\"{u}r Strahlen- und Kernphysik (HISK) for his contribution.
\end{acknowledgements}

\nolinenumbers

\end{document}